\documentclass[12pt]{article}
 \usepackage{longtable}
\relax
\usepackage{amssymb}

\newcommand{\bo}{{\bar o}}

\def\bo{{\raise.15ex\hbox{\large$\Box$}}}               

\def\face{{\raise.2ex\hbox{$\displaystyle \bigodot$}\mskip-2.2mu \llap {$\ddot
        \smile$}}}                                      

\def\ket#1{\left| #1\right\rangle}              
\def\leftrightarrowfill{$\mathsurround=0pt \mathord\leftarrow \mkern-6mu
        \cleaders\hbox{$\mkern-2mu \mathord- \mkern-2mu$}\hfill
        \mkern-6mu \mathord\rightarrow$}       
\def\dvec#1{\vbox{\ialign{##\crcr
        \leftrightarrowfill\crcr\noalign{\kern-1pt\nointerlineskip}
        $\hfil\displaystyle{#1}\hfil$\crcr}}}           

\def\beq{\begin{equation}}
\def\eeq{\end{equation}}

\def\beqx{\begin{displaymath}}
\def\eeqx{\end{displaymath}}

\def\beql{\begin{eqnarray}}
\def\eeql{\end{eqnarray}}

\newcommand{\bea}{\begin{eqnarray}}
\newcommand{\eea}{\end{eqnarray}}

\def\[{\left [}
\def\]{\right ]}
\def\({\left (}
\def\){\right )}

\def\+{\oplus}

\begin{document}

\vskip .5in

\begin{center}
{\Large \bf Schr\"odinger's Cat is not Alone }

\vspace*{.7in}

{Beatriz Gato\footnote{Gato is the Spanish word for cat. The author is a linear superposition of these two authors.} and Beatriz Gato-Rivera} \\

\vskip .5in

{\Large \bf Presented at April Fools' Day 2010}

\vskip .5in

\end{center}

\begin{center}
\vspace*{0.5in}
{\bf Abstract}
\end{center}
We introduce the `Complete Wave Function' and deduce
that all living beings, not just Schr\"odinger's cat, are actually described
by a superposition of `alive' and `dead' quantum states; otherwise they would never die. 
Therefore this proposal provides a quantum mechanical explanation to the world-wide
observation that we all pass away.  Next we consider the Measurement problem in the 
framework of M-theory. For this 
purpose, together with Schr\"odinger's cat we also place inside the box Rasputin's 
cat, which is unaffected by poison. We analyse the system identifying its excitations 
(catons and catinos) and we discuss its evolution: either to a classical fight or to a 
quantum entanglement. We also propose the $BSV\Psi$ scenario, which implements the
Complete Wave Function as well as the Big Bang and the String Landscape in a very
(super)natural way. Then we test the gravitational decoherence of the 
entangled system applying an experimental setting due to Galileo. We also discuss 
the Information Loss paradox. For this purpose we consider a massless black cat
falling inside a massive black hole. After that we outline a method to compute the contribution
of black cats to the dark matter of the universe.
Finally, in the spirit of Schr\"odinger, we propose that next generation double-slit experiments 
should use cats as projectiles. Cat interferometry will inevitably lead to the `Many Cats'
interpretation of Quantum Mechanics, allowing to shed new light on old mysteries and paradoxes.
For example, according to this interpretation, conservative estimates show that decision making 
of a single domestic cat will create about 550 billion whole universes every day, 
with as many replicas of itself. 

\vskip .4in

\noindent
April 2010

\newpage

{\ \ \ \ \ \ \ \ \ \ \ \ \ \ \ \ \ \ \ \ }

\vskip .5in

{\it $<<$ If one wishes to provoke a group of normally phlegmatic physicists into a state
of high animation -- indeed, in some cases strong emotion -- there are few tactics better
guaranteed to succeed than to introduce into the conversation the topic of the foundations
of quantum mechanics, and more specifically the quantum measurement problem. $>>$ }

\vskip .3in

 {\it $<<$ It seems to me that the many-worlds interpretation is nothing more than a verbal
placebo, which gives the superficial impression of solving the problem at the cost
of totally devaluing the concepts central to it, in particular the concept of `reality'.
When it is said that the `other worlds' of which we are not, and in principle never 
could be, conscious, are `equally real', it seems to me that the words have become
uprooted from the context which defines their meaning and have been allowed to
float freely in interstellar space, so to speak, devoid of any point of reference,
thereby becoming, in my view, quite literally meaningless.  I believe that our
descendants two hundred years from now will have difficulty understanding how
a distinguished group of scientists of the late twentieth century, albeit still a
minority, could ever for a moment have embraced a solution which is such
manifest philosophical nonsense. $>>$ } 

\vskip .5in

Anthony James Leggett, Nobel Prize in Physics 2003, in `The problems of physics' \cite{AJL}

\newpage

{\bf Introduction}

\vskip .3in

Since its formulation in the 1930s, the foundations of Quantum Mechanics provided an endless source
of intense and passionate debate and inspiration. Far from obsolete, the interpretation of some features of
Quantum Mechanics has become a most vivid and popular topic of scientific inquire in the last three
decades or so. There are reasons for that. The Copenhagen no-interpretation (as A.J. Leggett wisely calls it),  
widely accepted because the alternatives were even worse, left too many questions unanswered.
Whereas the microscopic world was astonishing strange, counterintuitive and non-deterministic by first principles,
the macroscopic world dwelled confortably in familiar territory relying on, to a great extend intuitive, 
deterministic laws. Measurement of the microscopic entities by a classical device would produce the 
wave function to collapse into one of the possible states, the likelihood of each possible state being given by
the corresponding coefficient in the wave function  (the square of it, rather).  But, why a macroscopical 
classical device had the magical power to collapse any wavefunction at all? And, where was the divisory line 
between macroscopic classical and microscopic quantum objects? Wouldn't be better to propose that the collapse 
is produced by the mind/consciousness of the observer? After all, the mind and especially the consciousness are
weird and mysterious enough from the physics viewpoint. These questions, known as the Measurement problem
for short, gave rise to a plethora of different interpretations of Quantum Mechanics over the years,   
although not as successful as the Copenhagen one\footnote{ A short panoramic view of 
different interpretations of Quantum Mechanics can be seen in \cite{IQM}.}.
A central issue in any of such interpretations has been the applicability 
of the quantum formalism to increasinly bigger systems, including cats, the whole universe we inhabit, 
and currently even the multiverse\footnote{ See for example \cite{PyS}.}.

In this talk we present our 
own interpretation of Quantum Mechanics, not radiacally but still quite different from the Copenhagen one,
based on the hypothesis of the `Complete 
Wave Function' for macrophysical as well as microphysical systems or objects. Essentially the idea is: i) that 
classical systems or devices do not exist since everthing is ruled by Quantum Mechanics (we are not the first 
ones to say that), and ii) that wave functions are created again inmediately after they are destroyed by the collapse.
In addition, we discuss the Measurement problem in the framework 
of M-theory and propose the $BSV\Psi$ scenario that implements in a very (super)natural way the Complete 
Wave Function as well as the Big Bang and the String Landscape. Finally, we also consider some other
related issues like the Information Loss paradox. 
 
As shown by the title, special attention will be paid to Schr\"odinger's cat, as well as to some of its relatives. 
First of all we must admit, in order to be honest, that we have never understood the paradox named after it. 
What is wrong with a cat being alive and dead at the same time? We don't even know exactly what life and 
death are, nor we understand the border between both! For example, whereas some humans are buried alive 
by mistake, some others swear they died but decided to come back to life, a few moments later,
because they had something important to do. Many others claim to remember past lives (in different
bodies, of course), not to mention the millions who assure they have been visited by deceased loved ones.
So, unless Science clarifies once and for all the differences between being
alive and being dead, it does not make much sense to claim that a living being cannot be alive and dead
simoultaneously, at least for a short time\footnote{In the meantime, it is interesting to notice that at least
50\% of the world's population believe in the existence of the soul for both animals and humans and in
reincarnation for all of them (often mapping animals, men and women into each other). 
As far as we know, only the religions based on the Bible (Islam, Christianity and 
Judaism) believe in the existence of the soul exclusively for humans, regarding women's souls as second class 
with respect to the first class men's souls, and in a unique terrestrial (or extraterrestrial)
existence for each life form.}.

On the other hand, only one of the possible quantum states in the superposition is realized in the real world, 
whereas the others just evaporate in the collapse of the wave function. So, the paradox becomes even less, 
unless one doesn't believe in the collapse of the wave function... but this is not our problem. 
Having said this now we start properly.

\vskip .6in

{\bf Complete Wave Function}

\vskip .3in

Let us introduce the `Complete Wave Function' (CWF) for macroscopic as well as microscopic physical 
systems. The idea is that Quantum Mechanics rules everything in our physical universe, although not 
necessarily rules our universe as a whole, or other universes or the `mind/consciousness' universe. 
The CWF can be regarded as a very long sum of partial wave functions corresponding to specific
time coordinates each. For example, the CWF describing the entire existence of a cat will read:

\begin{equation}
\Psi(cat) = \sum_{r=0,1...}^{n} \sum_{t=t_i}^{t_f}  \psi(cat,t,\Delta t)  +  P  +  M \  ,
\end{equation}

\vskip .1in

\noindent
where $ \psi(cat,t,\Delta t) $ is the partial wave function produced at time t and collapsing only a $\Delta t$
later, which is a linear(?) superposition of possible outcomes of the cat due to its interaction 
with itself or with the environment, the values $t_i$ and $t_f$ indicating the initial and final times of
the cat's life. Since cats have the reputation to have many lives\footnote{In Spain cats are supposed to
have seven lives, whereas in some other european countries and in the USA they are supposed to 
have nine.}, we also sum over the number of reincarnations $r$, with $n$ the total number of them.  
Finally, $P$ and $M$ are integration constants that encode our ignorance on other possible 
physical and metaphysical degrees of freedom of the cat, respectively. Observe that, whenever
the cat has a non-vanishing probability of dying, then $\psi(cat,t,\Delta t)$ will be a superposition
of $\ket{alive}$ and $\ket{dead}$ states. 

In general, the most intuitive way to think of the CWF is simply as a continuous creation and collapse
of the partial wave functions describing the system. To be precise, at a given time there is a wave function 
at work that collapses very quickly in order that one of the possible outcomes is actually realized in the real 
world. Shortly after, another wave function should come into existence; otherwise nothing else could 
happen to the system anymore. This wave function should also collapse for the same reason as 
before,....and so on. Therefore the emergence of classical reality for macrophysical systems; i.e. the 
classical world, simply results from the continuous, and very fast, creation and collapse of the partial wave
functions describing all that exist. The only difference for microphysical systems is that the collapse
of the wave functions takes much longer, what allows us to prepare or encounter systems in which
the collapse didn't happen yet.

One can wonder why and how the partial wave functions are continuously created with very short 
time lapses between them, at least for macrophysical systems. We don't know the answer, of course,
but we regard these questions at the same basic level as:  why do wave functions exist at all?
In our opinion, there is no guarantee that we terrestrial humans
will ever understand this process, in case it correctly describes our physical world,
nor even the very existence of wave functions\footnote{In our opinion Einstein 
was overly optimistic and naive with statements of the kind:
`The most amazing thing about the universe is that it can be understood'. Far from truth, A. Einstein
never understood Quantum Mechanics, that underlies all the physical laws in our universe and
that, according to R. Feynman, nobody can understand. Observe, however, that individuals from some
extraterrestrial advanced civilizations could in principle understand scientific concepts and ideas
beyond reach of the terrestrial humans, due to their superior brain capabilities. }.
As to why the partial wave functions collapse very fast for macroscopic systems and apparently
much slower for microphysical systems, this is just a minor complication to the already intricate
Measurement problem, and will be discussed later in the $BSV\Psi$ scenario.

An important implication of the CWF hypothesis is that physical systems, including living beings, can 
only experience states or events that are possible outputs of the corresponding partial wave functions. 
Decease of living beings cannot be an exception, as we mentioned in the case of the cat. As a consequence, 
a necessary but not sufficient condition for a living being to die is that the corresponding partial wave 
function just before the decease be a superposition of $\ket{alive}$ and $\ket{dead}$ states. The other way
around, one can regard the existence of superpositions of $\ket{alive}$ and $\ket{dead}$ states as the
ultimate reason explaining why all living beings pass away. To conclude: Schr\"odinger's cat is 
not alone.

\vskip .6in

{\bf Measurement Problem}

\vskip .3in

Let us consider the following gedanken experiment. Together with Schr\"odinger's cat we also place inside 
the box Rasputin's cat, which has been trained by its owner to become inmune to poison. Then we close
the sound-proof box tightly, making sure no noises will escape to the exterior. In addition, for double security
check, we carefully place earplugs inside our ears, since hearing any meowing or scratching would constitute 
an undesirable observation that would ruin the experiment.  We assume that Rasputin's cat does not lose 
his mind inside the box and that both cats are in good health (no heart attacks allowed during the experiment).

To start we analyse the system identifying its excitations: catons and their supersymmetric partners catinos.
We find that the spectrum of both catons and catinos are quantized in units of a universal feline constant f.  
 
Next we study the system's evolution, either to a classical fight or to a quantum entanglement.
Our results indicate that the system will evolve first into a classical fight with 
100\%  confidence level, but after a while there is an increasing probability that the system
becomes an entangled one. An important factor here seems to be the gender of the cats and the season: 
the entanglement is more probable if the cats belong to different genders and the experiment is performed in 
the heat season. We found that it is also possible a partial entanglement involving only the tails.

Now let us discuss the Measurement problem in this situation. The two cats are entangled with each other
and with the poison bottle, that has 50\% probability to release the poison inside the box. As a result,
Schr\"odinger's cat's wave function shows 50\% 
probability of becoming alive and 50\% probability of becoming dead. The question is: what determines
the collapse, or reduction, of the wave function, the cat turning into either 100\% alive or 100\% dead? 
Here are some answers, all of them in the framework of M-theory :

\begin{itemize}

\item{1. The Maximal Materialistic explanation: It is the release of the poison inside the box what causes 
the collapse of Schr\"odinger's cat together with its wave function into 100\% dead. Otherwise, the uncontaminated air
will cause the collapse of the wave function into 100\% alive. Therefore the air inside the box plays the
r\^ole of the classical measurement device in the Copenhagen no-interpretation.}

\item{2. The Middle Materialistic explanation: The action of thermal fluctuations from Rasputin's cat and from
the walls of the box produce environmental decoherence due to entanglement overdose. This is not a honest 
reduction of the wave function, however, it only looks like that.}

\item{3. The Minimal Materialistic explanation: The action of gravity causes the reduction of the wave function,
also known as `gravitational decoherence'. R. Penrose has contributed remarkably to this line of 
thought in his best-seller books \cite{PenENM}, \cite{PenShM} and \cite{Penrose}. We will come back to this issue.}

\item{4. The Mystery or Magic explanation: The collapse of the wave function is produced by a mysterious cause
that resembles magic, completely unknown for science. Since this cause is completely unkown we cannot 
possibly know whether we like it or not.}

\item{5. The Mind explanation: The mind/consciousness of an earthly observer (mind, for short) produces the reduction 
of the wave function. It could be the mind of Schr\"odinger's cat itself or the mind of 
Rasputin's cat or the mind of the human opening the box at the end of the experiment. 
The latter option we find too anthropic chauvinist, however, so we discard it.
In any case, if Schr\"odinger's cat dies because of the poison, Rasputin's cat will be the first living being with
a mind that will observe and notice the decease. In this case the observer's mind plays the r\^ole of the classical 
measuring device of the Copenhagen no-interpretation. This implies that the mind of any observer 
behaves in ways not predicted by, and even in contradiction with, Quantum Mechanics; 
that is, the latter does not rule the mind/consciousness `universe'.
E. Wigner supported this idea very enthusiastically in an article in 1961 \cite{Wigner}, whereas R. Penrose supports, very
enthusiastically too, that mind/consciousness has no place in Classical Physics \cite{PenENM}, \cite{PenShM}.}

\item{6. The Metaphysical explanation: The mind/consciousness of a heavenly observer produces the collapse of the wave 
function. For example, an angel-cat should always be around, and an evil demon-cat should be lurking in such circumstances
for in case he has to pick up the cat's soul (heavenly observers can read the probabilities of the possible 
outcomes of the earthly wave functions, so they prepare themselves for the occasion).}

\item{7. The Mystical or Miracle explanation. The collapse of the Schr\"odinger's cat's wave function 
is produced by a very powerful mind/consciousness that stretches along the whole universe. To this category belong 
the various versions of the Creator given by several religious and metaphysical views (not all religions
and metaphysical schools believe in a Creator, though). Following some tradition in Quantum Mechanics 
and Particle Physics, below we propose the $BSV\Psi$ model by invoking the Hindu Creator, i.e. the Trimurti: 
Brahma, Shiva and Vishnu\footnote{Because of this tradition, it doesn't come as a surprise to encounter 
at CERN, in between the hostel buildings, a sculpture of Shiva performing the `The Cosmic Dance'. }. 
As we will see, this scenario provides a very (super)natural explanation for the existence of the Complete 
Wave Function, as well as for the Big Bang and the String Landscape\footnote{The Christian Creator 
has also three manifestations, i.e. the Trinity: Father, Son and Holy Spirit. However, we don't know how to 
implement the Complete Wave Function, neither the Big Bang nor the String Landscape in that framework.}.}   

\end{itemize}

\vskip .6in

{\bf  BSV$\Psi$ Model}

\vskip .3in

According to the Hindu cosmology Brahma, the primordial Creator, produced the Big Bang using `sound technology'. 
Namely, Brahma, repeating or chanting in His Mind the `sounds of creation'  (such as Om, rather oooooooommmmmmm), 
with a specific combination of sounds
and rythm, created our universe with its specific laws of physics. After the Big Bang Brahma lost interest in 
the project and got busy preparing another Big Bang for another universe, with presumably different laws
of physics than ours. 
After that, Brahma again created a new universe and kept on creating more and more universes, the laws of 
physics of each universe (gauge groups, particle spectrum, cosmological constant, whether or not supersymmetry, etc.)
being determined, presumably, by the specific sounds and rythms, etc., used by Brahma's Mind. We see therefore that the
Hindu cosmology, which comes equipped with the Big Bang, also provides a very (super)natural explanation for the String 
Landscape. The other way around, the huge String Landscape explains why Brahma is always busy creating universes. 

After the Big Bang, Shiva and Vishnu took over the universe just created by Brahma: a mess of wave functions, in order
that it could come into real existence. Now we propose that the interplay between Shiva and Vishnu implements the Complete
Wave Function hypothesis. The major r\^ ole of Shiva, the Destroyer, is precisely to destroy; i.e. to collapse most wave functions 
so that Classical Physics and reality can emerge. Then, Vishnu, the Preserver or Sustainer, must create very quickly new wave 
functions for everything and everybody again, otherwise the universe would simply freeze and nothing will happen anymore.
Shiva destroys again most wave functions created by Vishnu, who promptly creates new ones... etc. As a result,
this scenario accounts for the continuous emergence of Classical Physics out of Quantum Physics: due to the interplay 
between Shiva and Vishnu, the classical universe exists and evolves with all its macrophysical objects, systems and 
creatures. In particular, the collapse of the Schr\"odinger's cat's wave function would have been produced by
Shiva's Mind.

Regarding the microworld, where Quantum Mechanics is more evident, either Shiva does not pay attention to
this realm, or He pays attention but wisely decides not to produce the collapse of the wavefunctions for such 
systems until an appropriate moment ......  to fool humans, for example.
Otherwise Classical Mechanics would take over at the microscale and we humans would have never discovered Quantum 
Mechanics (perhaps life couldn't even exist in such scenario).

\vskip .6in

{\bf Testing Gravitational Decoherence}

\vskip .3in

Now let us come back to the box. At this point we are ready to test the gravitational decoherence 
of the entangled system. For this purpose, in the spirit of Galileo we climb the tallest tower in town and 
throw the box from the top. Careful analysis of the remains will give us valuable information, not only
from the gravitational decoherence of the system, but also from other important issues such as the entanglement 
entropy versus total entropy (number of scratches on the cats skins versus number of 
scratches everwhere inside the box). In the most than probable case that Schr\"odinger's cat doesn't
survive these experiments, analysing the air composition inside the box will indicate whether the decease
was caused by the action of the poison or by the action of the gravitational decoherence\footnote{More tests of this 
kind are discussed in appendix 2 of \cite{PenLSM}:  `Experiments to test gravitationally induced state reduction'.}.
Rasputin's cat has many more chances to survive as it masters many shamanic tricks.

\vskip .6in

{\bf Information Loss Paradox}

\vskip .3in

First of all we do not understand all that noise and worrying about losing information. As a matter of fact, following the daily 
news it seems that it would be a blessing if more than 90 percent of the information of this planet would disappear forever.
Having said that we come to the problem at hand.

Let us consider a massless black cat falling inside a massive black hole. While heading towards the black hole at the speed of 
light, the cat, being massless and black, only feels the gravitational pull through its own black-body radiation and the quantum
information stored in its brain, in the lines described in references \cite{Saurav} and \cite{LKL}, respectively. Once at the 
event horizon, only the quantum information can go further, by tunneling, the rest of the cat and its black-body radiation
spreading all over the place, especially the hair. Now the qubits of information will reach the singularity or will 
evaporate converted in thermal Hawking radiation, adding to the cat's black-body radiation at the horizon. In both cases 
the information will be lost forever, it seems....... At least this will be the conclusion drawn by most materialistic readers. 
For metaphysical oriented readers, however, there are a bunch of more possibilities. First of all, a copy of the cat's mind  
will remain in the cat's soul. Second, there will be also copies in several detailed dossiers kept by angel-cats and 
demon-cats and the like, as well as by Vishnu and Shiva and the like. Moreover, another good quality copy will be
preserved forever in the Akashic Records\footnote{Eastern religions as well as many Western metaphysical schools assure 
that there exists a substance named by the Sanskrit word `Akasha' penetrating everywhere (similar to the 
non-vanishing v.e.v. of a scalar field, or a cosmological constant, or aether, in modern language).  It exists in all 
dimensions, i.e. in the branes and in the bulk, and records all that occurs, including thoughts, feelings and intentions of
living creatures. Skilled people are supposed to be able to retrieve any information whatsoever by reading the 
`Akashic Records' using appropriate techniques. See, for example, the popular metaphysics guide \cite{Rampa}.}.

The alert reader will point out that the major problem with loss of information is, however, that it shows (once
again) that Quantum Mechanics and General Relativity are incompatible. That is, either Quantum Mechanics is
not a complete theory or General Relativity is not the ultimate theory of gravitation. In our opinion, neither
Quantum Mechanics nor General Relativity are complete definitive theories, as deduced from the {\it Gato Conjecture 1}:

\vskip .1in

{\it Technological civilizations only start discovering the ultimate laws of physics 
after at least 400 years from the discovery and mastery of the electromagnetic waves}.

\vskip .1in

However, there are obstructions for the technological civilizations in their way to discover the definitive laws
of physics, as deduced from the {\it Gato Conjecture 2}:

\vskip .1in

{\it Technological civilizations only survive, in average, 200 years after the invention of TV due to  
neural system degeneracy resulting in generalized irreversible brain damage}.

\vskip .1in

Therefore very few civilizations in the universe will manage to write down an ultimate theory of anything at all.

\vskip .6in

{\bf Contribution of Black Cats to the Dark Matter of the Universe}

\vskip .3in

Now let us outline how to estimate the contribution of black cats to the dark matter of the universe. A honest 
computation from scratch should start with a Drake-like equation for feline civilizations\footnote{See ref.
\cite{Gleiser} for a recent interesting discussion on the Drake equation.}. However there is a shortcut. One 
has to take into account that cats can only proliferate when they have servants \cite{Kilin}. Therefore, 90\% 
of the cats in the universe will cluster in planets where there are human-like technological civilizations. As 
a result we only need to borrow the estimates of the SETI experts for such civilizations and make some suitable 
corrections and generalizations. Notice that the SETI estimates refer only to the amount of civilizations that at 
present could be able to release enough electromagnetic junk that we could observe. Therefore the SETI data 
refer only to the number of civilizations in our galaxy with a technological level equal or slightly superior than ours
(presumably, really superior civilizations do not contaminate anything anymore).  On the other hand, since the Milky 
Way is a typical barred spiral galaxy, we can always generalize the results to other galaxies, with some correction 
factors, and make some averages. Taking all these considerations into account, we find the following steps to be taken
to compute the total mass of black cats in the universe using the SETI estimates.

\begin{enumerate}
\item{One has to multiply by a factor of at least 10000 the SETI data in order to get the number of civilizations
in our galaxy that can host cats. The result will give us 90\% of the feline civilizations in our galaxy.}
\item{Multiply the number of feline civilizations in our galaxy by an average number of individuals in each
and by the fraction of them which are black.}
\item{Multiply the number obtained by an estimate average mass of each individual and we get the contribution
of black cats to the dark matter of the Milky Way.}
\item{Multiply this contribution by the number of galaxies similar to ours and by the number of other galaxies 
not so similar but with a correction factor.}
\end{enumerate}

The proponents of the "We are Alone" scenario disagree completely with the SETI estimates though. They 
rather prefer the {\it Strong Felinepic Principle} that reads:

\vskip .1in

{\it Feline civilizations, which are the superior life forms in the universe, only exist in planet Earth. Therefore 
extraterrestrial cats do not exist, and the whole universe has been created for the solely purpose of supporting
Feline life in planet Earth.}

\vskip .1in

Observe that this principle may compromise financial support for any future SETFI (Search for Extraterrestrial Feline 
Intelligence) project.

\newpage

{\bf Many Cats Interpretation} 

\vskip .3in

The double-slit experiment has been running for more than two centuries by now. As is well known, it was 
designed in 1801 by T. Young in order to investigate the wave nature of light. Even though Young's
results showing diffraction patterns were very relevant at that time, the importance of the experimental
setting increased enormously in the twenty century with the advent and development of Quantum Physics. 
For a very long time the experiment, and its variations, made use of photons exclusively. In 1961 for the
first time electrons were used as projectiles, followed some years later by neutrons.  The tendency to
apply the experiment to bigger, `more macroscopic' particles continued until the present, with the use 
of atoms and even some molecules.

Now we propose, in the spirit of Schr\"odinger, that next generation double-slit experiments should 
consider using cats\footnote{However, we do not recommend cat-cat collisions at the LHC at CERN, as the 
high production rate of black cats could endanger life on Earth.}. Following the logical steps in ref. \cite{Tag}
one finds that cat interferometry will inevitably lead to the `Many Cats' interpretation of Quantum Mechanics, 
allowing to shed new light on old mysteries and paradoxes. To get an idea of what this interpretation amounts,
we have made a follow up, during a whole ordinary day, of the decisions taken by a domestic cat\footnote{By 
domestic cat we mean a cat living in a house or a flat with domestic humans who believe they are 
the owners and pay all the bills. Notice that cats domesticated humans about 10.000 years ago \cite{Kilin}.}.
We found 39 decisions during the periods when the cat looked fully awake. As a result,
39 branches will be produced from this universe alone. However, each of these branched universes has a 
copy of the same cat and therefore will experience a number of branches itself. The new branched universes
will undergo some more branches in turn, etc. It is easy to see that the resulting total branching of 
universes due to the 39 decisions taken by the `original' cat, with identical copies of itself, is
$2^{39} - 1$, what amounts to about 550 billion, to be precise 549755813887, whole universes every day.

Taking into account that cats have an average life of 16 years (except in China), and that 16 years amount
to 5844 days, we conclude that a domestic cat will take about 227916 decisions during its lifetime,   
producing about $2^{227916}$ whole universes with as many replicas of itself.

We also speculate that the angular momentum of the newly created universes will be directly related to the tail's
helicity of the cat at the branching time, and that this helicity could influence some properties of elementary 
particle physics as well, like the chirality involved in week interactions, which in our universe is `left'.

\vskip .5in
\noindent
{\bf Acknowledgements:}
\vskip .2in
\noindent
We are very grateful to Kil\'\i n for patiently teaching us, over the years, all about status and relationships between 
humans and cats. Kil\'\i n, who was almost identical to Socks (the First Cat under the Clinton administration) 
was the best catfriend we have ever had. Certainly, without his influence this article would have never appeared. 
We also thank Chris Hull for the suggestion that we should collaborate with each other.

\bibliography{REFS}

\end{document}